\documentclass[preprint,twocolumn,showpacs,superscriptaddress,groupedaddress]{emulateapj} 

\usepackage{graphicx}  
\usepackage{dcolumn}   
\usepackage{bm}        
\usepackage{amssymb}   
\usepackage{tikz}
\usepackage{float}

\begin{document}

\title{Multi-messenger Extended Emission from the compact remnant in GW170817}



\author{Maurice~H.P.M. van Putten}
\affil{Sejong University, 98 Gunja-Dong Gwangin-gu, Seoul 143-6\%747, Korea; \href{Corresponding author.}{E-mail: mvp@sejong.ac.kr}} 
\author{Massimo~Della Valle}
\affil{Istituto Nazionale di Astrofisica, Osservatorio Astronomico di Capodimonte, Salita Moiariello 16, I-80131 Napoli, Italy} 
\author{Amir~Levinson}
\affil{School of Physics and Astronomy, Tel Aviv University, 69978 Tel Aviv, Israel}

\begin{abstract}
{GW170817/GRB170817A probably marks a double neutron star coalescence.
Extended Emission $t_s\simeq (0.67\pm0.03)$\,s post-merger shows an estimated energy output ${\cal E}\simeq (3.5\pm1)\%M_\odot c^2$ determined by response curves to power-law signal injections, 
where $c$ is the velocity of light. It provides calorimetric
evidence for a rotating black hole of $\sim 3M_\odot$, inheriting the angular momentum $J$ of the merged 
hyper-massive neutron star in the immediate aftermath of GW170817 following core-collapse about or 
prior to $t_s$. Core-collapse greatly increases the central energy reservoir to $E_J\lesssim 1M_\odot c^2$, 
accounting for ${\cal E}$ even at modest efficiencies in radiating gravitational waves through a non-axisymmetric thick torus. 
The associated multi-messenger output in ultra-relativistic outflows and sub-relativistic mass-ejecta is consistent with 
observational constraints from the GRB-afterglow emission of GRB170817A and accompanying kilonova.}
\end{abstract}


\nopagebreak

\section{Introduction}
GW170817 \citep{abb17a,abb17b} is the first observation of a low-mass compact binary coalescence seen in a long duration 
ascending gravitational-wave chirp. By the accompanying GRB170817A identified by 
{\em Fermi}-GBM and INTEGRAL \citep{con17,sav17,gol17,poz18,kas17a} it represents the merger of a neutron star with another
neutron star (NS-NS) or companion hole (NS-BH) with a chirp mass of about one solar mass. Potentially broad 
implications of the first has received considerable attention for our understanding of the origin of heavy 
elements \citep{kas17b,sma17,pia17,dav17}  and for entirely novel measurements of the Hubble constant \citep{gui17,fre17}. 

By chirp mass, the nature of GW170817 is inconclusive in the absence of observing final
coalescence at high gravitational-wave frequencies \citep{cou19}. For NS-NS coalescence, numerical simulations \citep[e.g.][]{bai17}  
show gravitational radiation to effectively satisfy the canonical model signal of binary coalescence in a run-up to about 
1\,kHz, beyond which the amplitude levels off and ultimately decays as the two stars merge into a single object at a 
maximal frequency $\sim$3\,kHz. In contrast, NS-BH mergers include  
tidal break-up \citep{lat76}. In a brief epoch of hyper-accretion, the black hole would be near-extremal with a remnant
of NS debris to form a torus outside its Inner Most Stable Circular Orbit (ISCO). This process is marked 
by gravitational radiation switching off early on at a frequency 500-1500\,Hz \citep{val00,fab09,eti09,fer10} 
and possibly quasi-normal mode oscillations at yet higher frequencies \citep[e.g][]{yan18}. 

Here, we report on the energy output ${\cal E}$ in gravitational radiation post-merger, that appears as a descending chirp of 
Extended Emission marking spin-down of a compact remnant to binary coalescence at a Gaussian equivalent 
level of confidence of 4.2$\sigma$ \citep{van19}. We give a robust estimate of ${\cal E}$ by response curves determined by 
signal injection experiments in data of the LIGO detectors at Hanford (H1) and Livingston (L1). 
${\cal E}$ introduces a new calorimetric constraint that may break the degeneracy of a NS or BH central engine.

${\cal E}$ reported here points to core-collapse of the merged NS produced in 
GW170817, inheriting its angular momentum $J$ while greatly increasing the associated
spin-energy $E_J$ through collapse to a Kerr BH \citep{ker63}. 

After our injection experiments were initiated, we learned of an independent analysis of energy considerations by single-template 
injections, pointing qualitatively to similar energies without, however, identifying the origin of our Extended Emission \citep{oli19}. 


\section{${\cal E}$ from pipeline response curves}

We set out to determine response curves of our search pipeline by signal injections into LIGO data (\cite{val14}; Figs. 1-2),
including whitening, butterfly filtering and image analysis of merged (H1,L1)-spectrograms (Appendix). Whitening is by normalizing the 
Fourier spectrum over an intermediate bandwidth of 2\,Hz, bringing about GW170817 more clearly than without whitening 
\citep{van19}. 

\begin{figure}
\vskip-0.35in
\centerline{\includegraphics[scale=0.28]{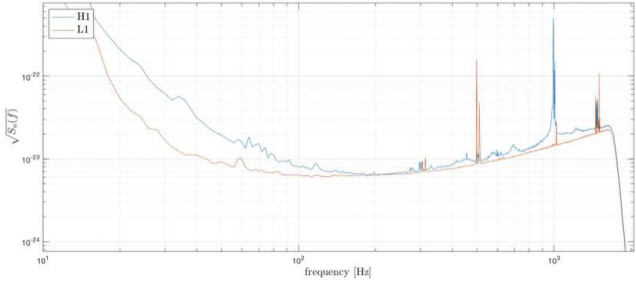}}  
\vskip-0.5in
\caption{H1 and L1 detector noise shown by the square root of spectral energy density $S_n(f)$ (at reduced
sampling rate 4096Hz with glitch in L1 removed by LIGO) for an epoch of 2048 s containing GW170817 (top panel). 
Spikes are violin modes associated with suspension of optics. Frequencies up to about 1700\,Hz can be used in
injection experiments. H1 and L1 detector noise is very similar during GW170817.}
\end{figure}

We recall that GW170817 is observed as an {\em ascending chirp} signifying the merger of two compact stars with time-of-coalescence 
$t_c=1842.43$\,s followed by GRB170817A across a gap of 1.7\,s. In our injection experiments to LIGO data, we include 
a model DNS with the same chirp mass ${\cal M}_c=1.188M_\odot$ of GW170817 (Fig. 2) at time-of-coalescence about 1818\,s,
producing two ascending chirps side-by-side (Fig. 3). A DNS is described by binary masses $M_1$, $M_2$, $\mu = M_1M_2/M$, 
$M=M_1+M_2$, at orbital separation $a$ and orbital frequency $\Omega \simeq c\sqrt{R_g/a^3}$ $(a>>R_g$), where $R_g=GM/c^2$ 
is the gravitational radius of the system, given the velocity of light $c$ and Newton's constant $G$. This merger chirp has a 
quadrupole gravitational-wave frequency $f_{GW} = \pi^{-1}\Omega$, 
\begin{eqnarray}
f_{GW}(t)= A(t_c -t )^{-\frac{3}{8}}~~(t<t_c),
\label{EQN_fgw1}
\end{eqnarray}
$[A]$\,=\,s$^{-5/8}$\,Hz, with strain $h(t) = ({4\mu}/{D})\left(M\Omega \right)^\frac{2}{3}$, $h(t) \simeq 1.7 \times 10^{-22} \left({M}/{3M_\odot}\right) \left({D}/{40\,\mbox{Mpc}}\right)^{-1} \left({f_{GW}}/{250\,\mbox{Hz}} \right)^\frac{2}{3}$ and $L_{GW} = ({32}/{5})\left({\cal M}_c\Omega\right)^{{10}/{3}}L_0$, where $L_0=c^5/G\simeq 200,000 M_\odot c^2$s$^{-1}$ \citep[e.g.][]{fer10}. 
For GW170817, $A\simeq 138\,$s$^{-5/8}$\,Hz. 
Up to 260\,Hz in both H1 and L1, $L_{GW}$ reaches $1.35\times 10^{50}\mbox{erg\,s}^{-1} \simeq 7.5\times 10^{-5}M_\odot c^2\,\mbox{s}^{-1}$, i.e., $4\times10^{-10}L_0$. 
While small compared to $10^{-5}L_0$ of GW150914 at similar frequency, GW170817 produced the largest strain observed by its 
proximity of $D\simeq 40$Mpc. It emitted $E_0=0.43\%M_\odot c^2$ over 200-300\,Hz with $h=1.4-1.8\times 10^{-22}$ over $\Delta t \simeq 0.25\,$s across $74\,\mbox{km} < r < 97 \,\mbox{km}$ assuming $M_1=M_2$.

A merged (H1,L1)-spectrogram shows Extended Emission post-merger below 700\,Hz in the form of an exponential feature
\begin{eqnarray}
f_{GW}(t)=(f_s-f_0)e^{-(t-t_s)/\tau_s}+f_0~~(t>t_s)
\label{EQN_fgw2}
\end{eqnarray} 
with the observed $\tau_s=3.01\pm0.2\,$s, $t_s=1843.1$\,s, $f_s=650\,$Hz and $f_0=98$\,Hz. 
For illustrative purposes, we note the isotropic equivalent strain $h=L_{GW}^{1/2}/(\Omega D)$ (in geometrical units, $c=G=1$) for the chirp mass of 
a small quadrupole mass-moment $\zeta=\delta m/M$ gives
$h(t) \simeq 2.7 \times 10^{-23} \left({\zeta}/{3\%}\right) \left({D}/{40\,\mbox{Mpc}}\right)^{-1} \left({f_{GW}}/{650\,\mbox{Hz}} \right)^{{2}/{3}}$, $L_{GW} \simeq$ $2\times 10^{52} \left({\zeta}/{3\%}\right)^2 \left({10M}/{r} \right)^5 \mbox{erg~s}^{-1}$ $\simeq 1\% M_\odot c^2/$s$^{-1}$. 

We next use phase coherent injections with frequency evolution (\ref{EQN_fgw2}) (Supplementary Data). 
No change in results below are found upon including phase incoherence by a Poisson distribution of random phase jumps over 
intermediate time scales, down to the duration $\tau = 0.5$\,s of our butterfly templates. 
$\tau=0.5$ appears intrinsic, as the Extended Emission feature tends to fade out as $\tau$ approaches 1\,s.

\begin{figure}
\hskip0.12in\centerline{\includegraphics[scale=0.28]{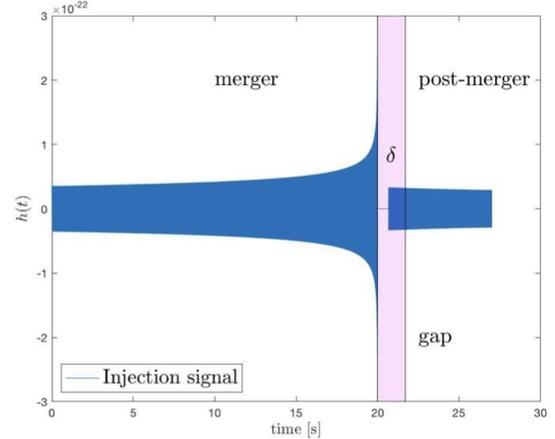}}  
\centerline{\includegraphics[scale=0.3]{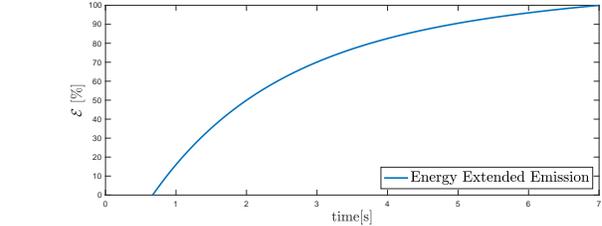}}  
\caption{Injection signal comprising a DNS merger and a post-merger branch separated by a delay $\delta = 0.67$\,s inside 
the gap of 1.7\,s between GW170817 and GRB170817A. The post-merger signal has a duration of 7\,s at with relatively
flat strain $h\propto f^\alpha$ ($\alpha=0.1$).}
\end{figure}

The total energy output ${\cal E} =\int_0^T L_{GW}dt$ is computed numerically as sums ${\cal E}=E_0 K^{-1}\sum \nu_i^2h_i^2$ (samples at $t_i$, $i=1,2,..,n$) covering a post-merger interval of duration $T$, where $K=\sum \nu_{0,j}^2h_{0,j}^2$ (samples at $t_j$, $j=1,2,\cdots, m$) is a 
reference sum with energy $E_0$ over a duration $T_0$, where $\nu_i=f_{GW}(t_i)$ denotes gravitational-wave frequency. 
$E_0$ is conform quadrupole emission in the {same orientation} of the progenitor binary by conservation of orbital-to-spin
angular momentum in transition to its remnant. Blind to any model in particular, we consider injections with power-law 
strain $h\propto f^\alpha$ with $T=7$\,s. 

Fig. 3 shows the outcome of a signal-injection alongside GW170817EE after a calibration $C_h=0.7$ for observed-to-true strain 
due to non-ideal H1 and L1 detector orientations relative to GW170817. Extended to multiple injections, results show there is 
no interference with the merger signal or with one another. 

Fig. 4 shows our estimated response curves $\chi({\cal E})$ for $h\propto f^\alpha$ ($0.1\le\alpha\le 1.0$). By $\hat{\chi}\simeq7.2$ of the Extended Emission to GW170817, we infer
\begin{eqnarray}
{\cal E}\simeq (3.5\pm1)\%M_\odot c^2,
\label{EQN_E}
\end{eqnarray}
For the descending chirp at hand (\ref{EQN_fgw2}), ${\cal E}$ mostly derives early on at high $f_{GW}$ with $L_{GW}\lesssim 1\%M_\odot c^2$s$^{-1}$ (Fig. 2). 

\section{Enhanced $E_J$ in collapse to a black hole}

${\cal E}$ in (\ref{EQN_E}) is a significant amount of energy, exceeding the merger output observed up to about 300\,Hz, emitted
as a descending chirp over a secular time scale of seconds with $f_{GW}<700\,$Hz far below the characteristic 
frequency $c/R_S\simeq 30$\,kHz of the Schwarzschild radius $R_S=2R_g$. Important energies also appear in 
GRB170817A and mass-ejecta \citep[e.g.][]{moo18a,moo18b}. Of these, ${\cal E}$ and $f_{GW}$ will serve as primary 
observational constraints on the remnant, i.e., $E_J$ of a rapidly spinning merged NS or rotating BH. 

\begin{figure*}
\centerline{\includegraphics[scale=0.39]{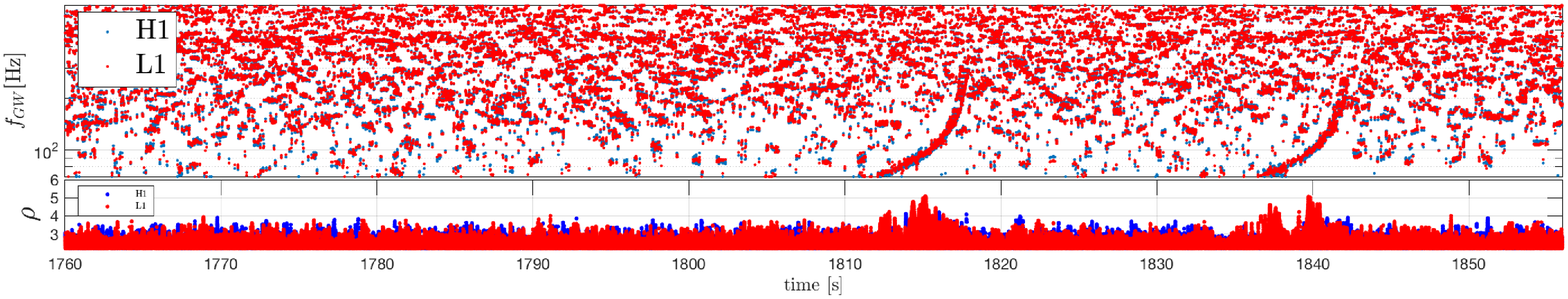}}  
\centerline{\includegraphics[scale=0.347]{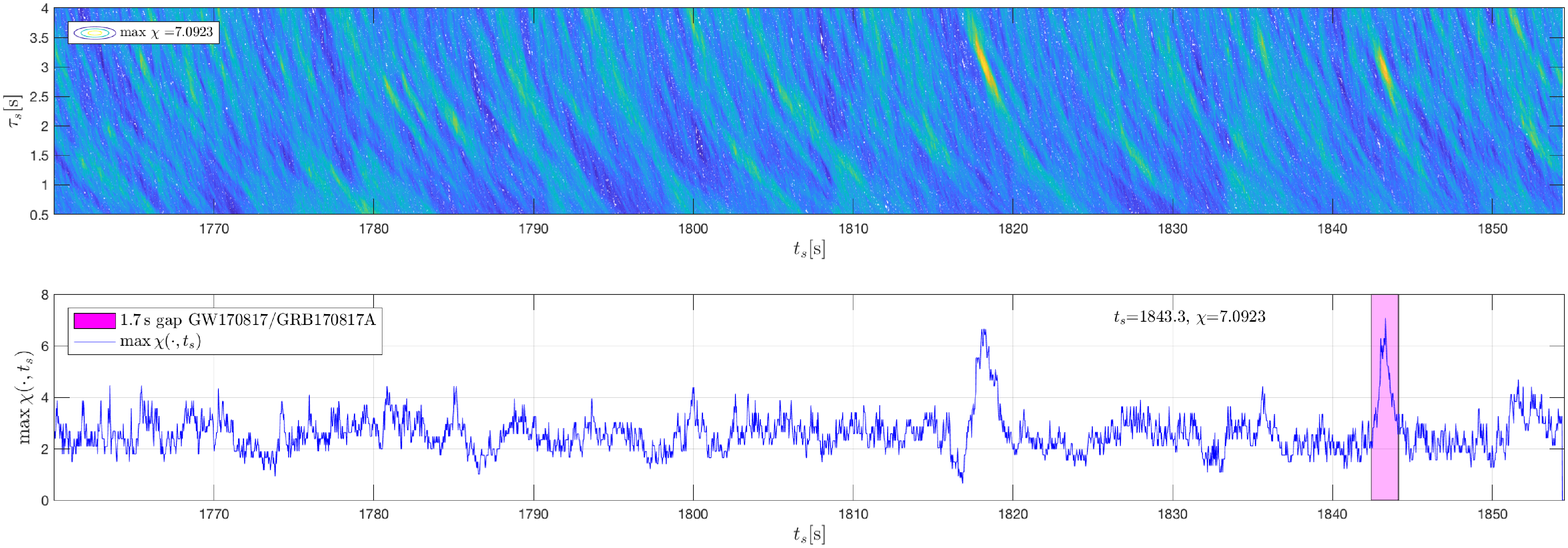}\hskip0.2in}  
\caption{(Upper panel.) (H1,L1)-spectrogram merged by frequency coincidences of butterfly filtering showing 
GW170817 ($t_c=1842.43$\,s) alongside a model signal ($t_c\simeq 1818\,$s). GW170817 appears with 
${\cal E}\simeq 3\%M_\odot c^2$ in Extended Emission. Included is $\rho=\sqrt{\mbox{SNR}}$ in the tail $>2$ 
of butterfly output of H1 and L1. (Lower panel.) $\chi$-image analysis of Extended Emission in the (H1,L1)-spectrogram over parameters $(t_s,\tau_a)$ for initial and final frequencies $(f_s,f_0)=(650,98)$\,Hz (upper panel) with similar signal strength of Extended Emission of signal injection to that of GW170817 ($t_s=0.67(\pm0.03)$s) measured by peak-value of the indicator $\chi$.}
\label{fig:ms}
\label{fig:chi}
\end{figure*}

While a long-lived NS might be luminous in gravitational radiation through a baryon-loaded 
magnetosphere (Appendix), its spin-frequency $f_s=(1/2)f_{GW}$ inferred from our Extended Emission is less than one-fifth
the break-up spin-frequency of about 2\,kHz. This modest initial spin limits $E_J$ to below $0.5\%M_\odot c^2$ and
probably somewhat less based on more stringent limits \citep[e.g.][]{hae09,oli19}. 

However, $E_J$ greatly increases by core-collapse of the merged NS in the immediate aftermath of GW170817, here at
time of core-collapse about or prior to $t_s=0.67(\pm0.03)$\,s post-merger (Fig. \ref{fig:chi}), where the 30ms
refers to our time-step in $t_s$. 
By the \cite{ker63} metric, 
\begin{eqnarray}
E_J = 2Mc^2\sin^2\left( \lambda/4\right) \lesssim 1M_\odot c^2 \left({M}/{3M_\odot}\right)
\end{eqnarray}
in terms of $a/M=\sin\lambda$, $J=a\sin\lambda$. This potentially
enormous energy reservoir amply accounts for ${\cal E}$ even at modest efficiency $\eta$, provided
a mechanism is in place to tap and convert $E_J$ into gravitational radiation. Moderate frequencies 
$f_{GW}<700\,$Hz can be realized in catalytic conversion into quadrupole emission by a non-axisymmetric disk or torus, 
sufficiently wide or geometrically thick. Exhausting $E_J$, a descending chirp results due to expansion 
of the ISCO during black hole spin-down.

\section{${\cal E}$ estimate from black hole spin-down}

To add some concreteness, we estimate $\eta$ in spin-down of an initially rapidly rotating BH, losing 
$J$ to matter in Alfv\'en waves through an inner torus magnetosphere \citep{van99,van01}. By heating, 
a non-axisymmetric thick torus is expected to generate frequencies correlated to but below those of a thin torus about 
the ISCO \citep{cow02}. In geometrical units, an extended torus produces emission from an 
orbital radius $r\equiv zR_g$ at twice the local orbital frequency, i.e., $f_{GW}\simeq c\pi^{-1}\sqrt{R_g/r^3}$. 
Asymptotic scaling relations for large radii (modest $\eta$) \citep{van03} show $L_{GW} \sim 10^{52} \mbox{erg~s}^{-1}$ 
for a non-axisymmetric torus with mass ratio $\sigma = M_T/M \simeq 0.1$. Accompanying minor output is in
MeV-neutrinos and $E_w\simeq \eta^2E_J$ in magnetic winds \citep{van02b,van03} - {\em most of $E_J$ is 
dissipated unseen in the event horizon, increasing area by Bekenstein-Hawking entropy \citep{van15}.} 
The observed 150\,Hz$ < f_{GW}<700\,$Hz indicates an effective radius of a quadrupole mass 
moment initially about three times the ISCO radius (Fig. 5), indicating a relatively thick torus. 
$f_{GW}$ decreases with $z$ with expansion of the ISCO during black hole spin-down. 

By numerical integration of this spin-down process, catalytic conversion of $E_J$ gives (Fig. 5)
\begin{eqnarray}
{\cal E} \simeq \left< \eta\right> E_{J} \simeq 3.6-4.3\% M_\odot c^2
\label{EQN_E2}
\end{eqnarray}
for canonical values of initial $a/M$, depending somewhat on the start frequency $f_s=600-700$\,Hz, 
consistent with (\ref{EQN_E}) inferred from $\chi({\cal E})$. 

The model estimate (\ref{EQN_E2}) uses  effective values of disk mass $m$ and $K$ throughout.
This does not readily predict $h(f_{GW})$ or the observed exponential feature (\ref{EQN_fgw2}), as
$m$ and $K$ will be time-dependent and vary with $z$. Use of effective mean values is only for 
our present focus on total energy output.

\begin{figure*}
\centerline{\includegraphics[scale=0.47]{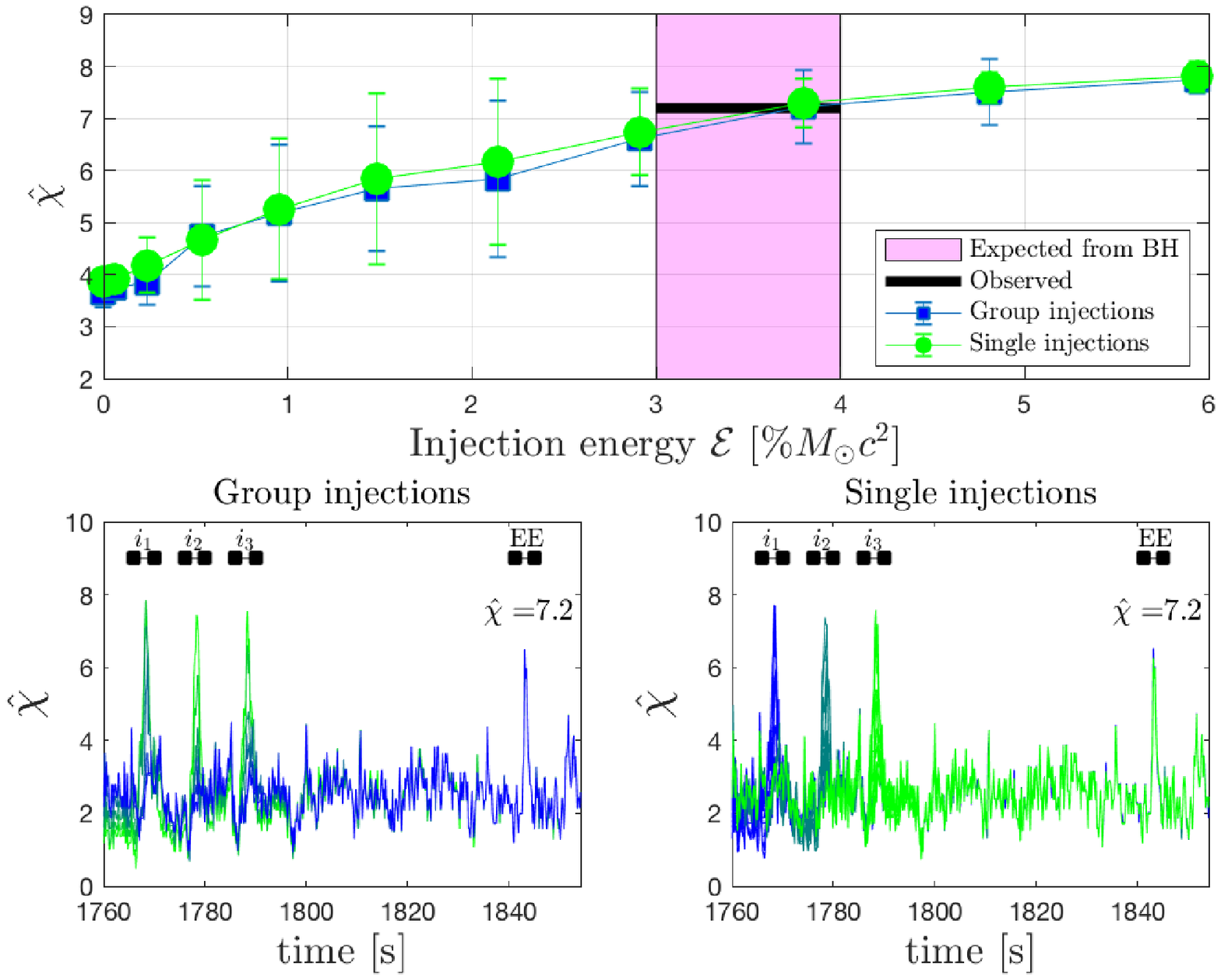}\includegraphics[scale=0.47]{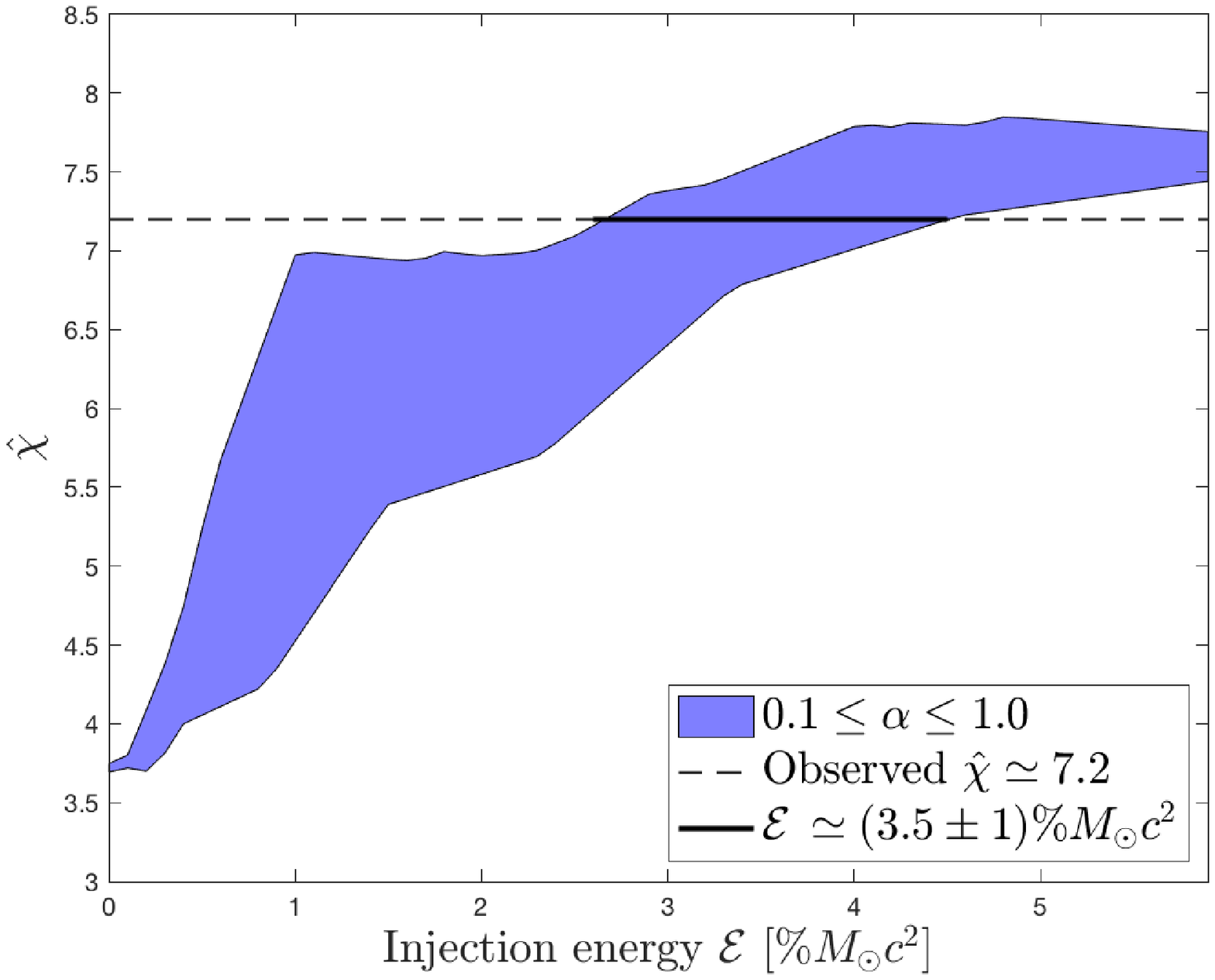}}
\caption{(Left panel.) Response peaks $\hat{\chi}({\cal E})$ of signal injections $h\propto f^\alpha$ ($\alpha=0.1$) 
in a merged (H1,L1)-spectrogram covering GW170817 by the indicator function $\hat{\chi}$ in $\chi-$image analysis as a 
function of energy input ${\cal E}$. The response curve (green, blue curves; top panel) determined by injections at
instances $(i_1,i_2,i_3)$ about 1 min before GW170817 (lower panels) is the same by grouped (left bottom) or single 
(right bottom) injections, demonstrating non-interference between different signals. 
Color (blue to green) indicates injection strength in group injections and injection position in single injections.
Scatter in $\chi(E)$ by noise fluctuations
appears least at $i_1$. (Right panel.) Extended Emission to GW170817 is observed at $\hat{\chi}\simeq7.2$ (blue dashed) intersected by
$\hat{\chi}({\cal E})$ from $i_1$ to power-law injections $h\propto f^\alpha$ ($\alpha=0.1,0.2,\cdots 1.0$) (blue filled).
}
\end{figure*}

\begin{figure}
\centerline{
\includegraphics[scale=0.49]{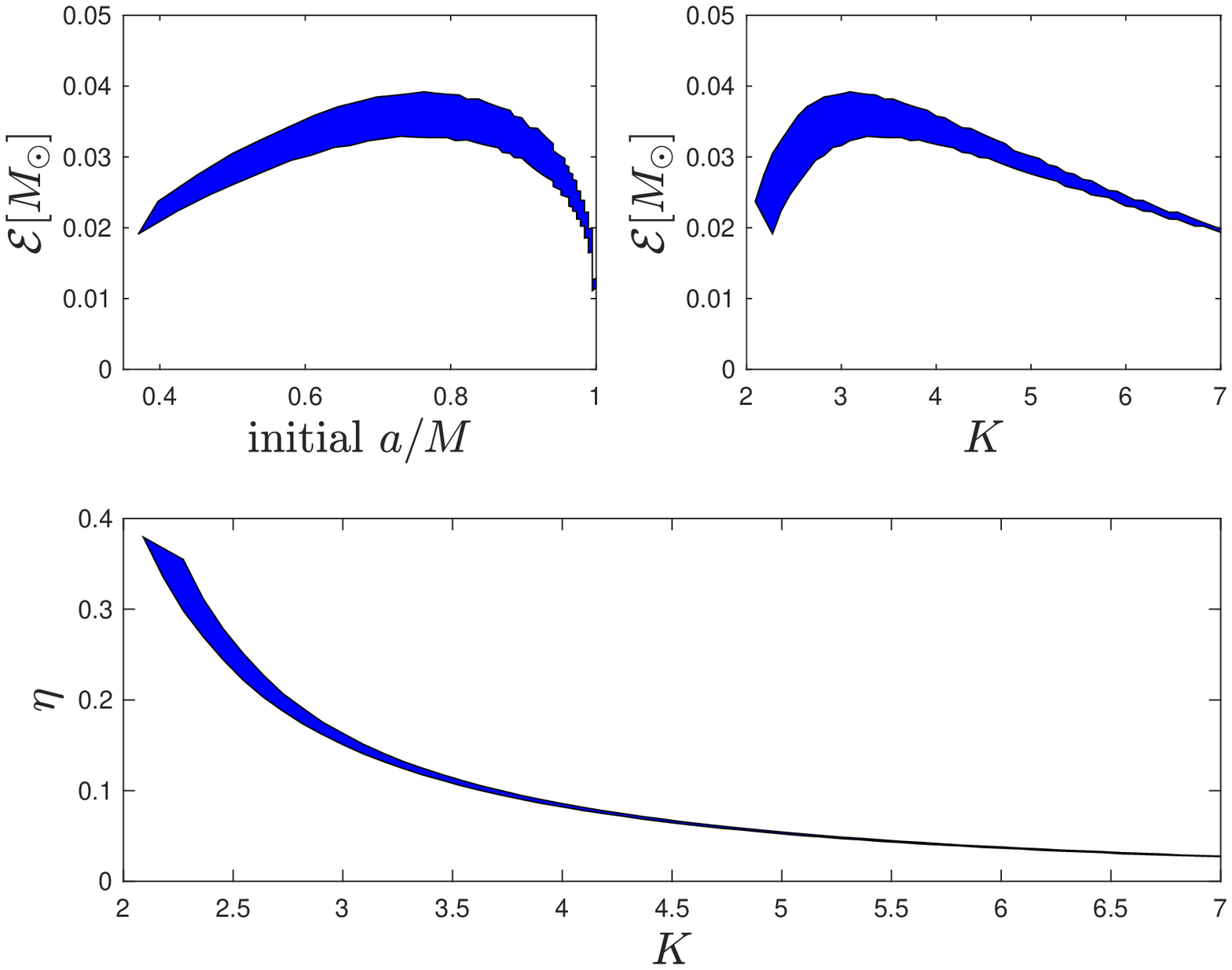}}
\caption{(Model prediction of ${\cal E}$ in a descending chirp from a non-axisymmetric torus of effective radius $K$ 
times the ISCO radius around a black hole of initial mass $M_0=3M_\odot$, converting $E_J$ into gravitational 
radiation at moderate efficiencies $\eta$. The boundaries of the thick curve refers to 
gravitational-wave frequencies $f_s=600-700$Hz at start-time $t_s$.}
\end{figure}

While the nature of GW170817 by the chirp up to 300\,Hz is somewhat inconclusive \citep{cou19},
${\cal E}$ provides a novel calorimetric constraint on its remnant. 
${\cal E}$ in (\ref{EQN_E}) challenges a hyper-massive NS \citep{oli19} yet is naturally accommodated by
(\ref{EQN_E2}) in core-collapse to a Kerr BH. 

In converting $E_J$, ${\cal E}$ is accompanied by MeV-neutrinos and magnetic winds \citep{van03} consistent 
with evidence for black hole spin-down in normalized light curves of long GRBs \citep{van12}. 

\section{Multi-messenger Extended Emission}

Starting with the merged NS from a DNS, a time-of-collapse about or prior to $t_s\simeq 0.67(\pm0.03)$\,s (Fig. 3) 
appears consistent - perhaps in mild tension - with the recently estimated time-of-collapse $0.98^{+0.31}_{-0.26}$\,s 
based on jet propagation times and mass of blue-ejecta \citep{gil19}.

Sustained by Alfv\'en waves outwards over an inner torus magnetosphere, a torus developing a dynamo 
with magnetic field $B=O\left( 10^{16}\right)$G limited by dynamical stability over the lifetime of black hole spin 
\citep{van03} gives a characteristic time scale for the lifetime of rapid spin of the BH and hence of the
BH-torus system, 
\begin{eqnarray}
T_s \simeq  1.5\,\mbox{s} \left( \frac{\sigma}{0.1}\right)^{-1}  \left( \frac{z}{6}\right)^{4}\left( \frac{ M}{3M_\odot} \right),
\end{eqnarray}
consistent with the duration $T_{90}$ (90\% of gamma-ray counts over background) of GRB170817A. 
Over this secular time scale, the BH gently relaxes towards a nearly 
Schwarzschild BH as the ISCO expands. A relatively baryon-poor environment of the BH is ideally suited for it to 
also launch an ultra-relativistic baryon-poor jet within a baryon-rich disk or torus wind with \citep{van03} 
\begin{eqnarray}
E_j \simeq \frac{E_J}{4z^4}\simeq 5\times 10^{50}\mbox{erg}, E_w \simeq \eta^2 E_J \simeq 4\times 10^{51}\mbox{erg},
\label{EQN_MM}
\end{eqnarray}
consistent with $E_j \sim 10^{49-50}$erg and $E_k = (1/2) M_{ej} v^2 \simeq 4.5\times 10^{51}$ in the relativistic ejecta of 
GRB170817A and $M_{ej}\simeq 5\%M_\odot$ of mass ejecta at mildly relativistic velocities $v\simeq 0.3c$ \citep{moo18a,moo18b}.
Emission terminates abruptly as the remnant torus collapses onto the black hole when $\Omega_H\simeq \Omega_T$ 
($f_{GW} \simeq 10^2\,$Hz).

\section{Conclusions}
${\cal E}\simeq (3.5\pm1)\%M_\odot c^2$ in Extended Emission measured by $\hat{\chi}({\cal E})$ through signal injections (Fig. 4) gives a powerful calorimetric constraint on the central engine. This outcome points to a Kerr BH formed in core-collapse of the merged NS in the immediate aftermath of GW170817. 

At $f_{GW}<700$Hz, our ${\cal E}$ is consistent with post-merger bounds of LIGO
\citep[][between dashed lines $E_{gw}=0.01-0.1M_\odot c^2$ in Fig. 1]{abb17c} and \cite{oli19}.
With $E_J$ of a Kerr BH, concerns of \cite{oli19} on detectability of Extended Emission are unfounded.
Accurate time-integration of the complex scaling $L_{GW}\propto\left(f_{GW} [h/C_h]\right)^2$ highlights 
a need for measurement by signal injection, for which a one-frequency estimate of $h_{H1}$ 
alone \citep{van19} now appears inadequate.

Core-collapse greatly enhances $E_J$ in $J$ inherited from the merged NS up to about $1M_\odot c^2$ in
a $\sim 3M_\odot$ BH. It amply accommodates ${\cal E}$ even at modest efficiencies in conversion to ${\cal E}$ 
over durations of seconds (Fig. 5). Accompanying minor emissions (\ref{EQN_MM}) in mass ejecta from the torus 
and ultra-high energy emission from the BH agree quantitatively with observational constraints on the associated 
kilonova and GRB170817A. GW170817 is too distant, however, to probe any MeV-neutrino emission \citep{bay12} 
its MeV-torus \citep{van03}. 

Conceivably, EE does not completely exhaust $E_J$, permitting low-luminosity latent emission including minor 
output in baryon-loaded disk winds and low-luminosity jets. While outside the present scope, 
such might be an alternative to the same from a long-lived NS remnant needed to account for 
ATo2017gfo \citep{ai18,li18,yu18,pir19}.

At improved sensitivity, LIGO-Virgo O3 observations may significantly improve on our ability to identify the nature of 
binary mergers involving a NS - including the tidal break-up in a NS-BH merger - and their remnants that might also be 
found in core-collapse supernovae and, possibly,  accretion induced collapse of white dwarfs.


{\bf Acknowledgements.} The authors thank the reviewer for a detailed reading and constructive comments.  
The first author gratefully thanks ACP, Aspen, Co, GWPop 2019 (PHY-1607611), and AEI, Hannover, where 
our signal injections were initiated in discussions with M. Alessandra Papa and B. Allen. We also thank A., V. Mukhanov 
and J. Kanner for constructive comments. We acknowledge use of the data set 10.7935/K5B8566F of the LIGO Laboratory 
and LIGO Scientific Collaboration, funded by the U.S. NSF, and support from NRF Korea (2015R1D1A1A01059793, 
2016R1A5A1013277, 2018044640) and MEXT, JSPS Leading-edge Research Infrastructure Program, JSPS Grant-in-Aid for Specially 
Promoted Research 26000005, MEXT Grant-in-Aid for Scientific Research on Innovative Areas 24103005, JSPS Core-to-Core Program, 
Advanced Research Networks, and ICRR.\\
\mbox{}\\
{\bf Supporting Data:}\\
\\
{\bf WInjection.m}, whitening and signal injection (Fig. 2), DOI 10.5281/zenodo.2613112\\
{\bf EEE.m}, estimated energy and efficiency of Extended Emission (Fig. 5), DOI 10.5281/zenodo.2613105\\

\appendix

Our broadband extended gravitational-wave emission (BEGE) pipeline aims for un-modeled ascending and descending chirps with 
a choice of intermediate time-scale of phase-coherence $0<\tau\lesssim1$s,
expected from extreme transient events exhausting $E_J$ of their central engine in seconds:
\begin{itemize}
\item Butterfly filtering is matched filtering against a bank of time-symmetric chirp-like templates of intermediate duration $\tau$, densely covering
a domain in $(f(t),\left| df(t)/dt\right|\ge\delta$) for some choice of $\delta>0$. Single detector spectrograms are extracted as scatter plots of correlations $\rho(t,f_c)$ between data segments (here, of 32 s duration) and time-symmetric chirp-like templates with central frequency $f_c$. 
\item To reduce noise in deep searches ($\kappa = 2$), spectrograms are merged by frequency coincidences ($\left| f_{c,H1}-f_{c,H2}\right| < \Delta f$) conform causality: $\Delta f$ is about $\left| df(t)/dt\right| \delta t$, where $\delta t=10$ms is the (maximal) signal propagation time between H1 and L1. We obtain satisfactory results with $\Delta f = 10$Hz (Fig. \ref{fig:ms}). 
\item Candidate features (Fig. \ref{fig:chi}) are evaluated by counting `hits:' $\chi$($\rho>\kappa\sigma$) by H1\&L1 over strips about a given family of curves - normalized to $\hat{\chi}$. For Extended Emission feature to GW170817, we use (\ref{EQN_fgw2}), giving $\hat{\chi}(t_s,f_s,f_0,\tau_s)$. The strip is of finite width ($\Delta f = 10$Hz, $\Delta t = 0.1$s), discretized with $\Delta t_s=0.030$s and, for background statistics, over $N = 16$ steps in each parameter gathered from 1956s of clean LIGO data in a scan over a total of 256M parameters ($N^3 = 4096$, $\Delta t_s=$30ms, \cite{van19}).
\end{itemize}

The merged NS produced by GW170817 may briefly emit GWs through a magnetosphere with field $B$, baryon-loaded with $M_b$ by dynamical mass ejecta and MeV-neutrino winds \citep[e.g.][]{per14}, by a quadrupole moment $\mu$ along its magnetic spin-axis misaligned with $J$ \citep{kal12}, extending out to $l$ of its light cylinder. At Alfv\'en velocity $c_A = B/\sqrt{ 4\pi \rho}$, $B=B_{16}10^{16}$G with matter density $\rho$, $\mu$ greatly exceeds that of $B$ in vacuum \citep{hac17}. In geometrical units ($c=G=1$), the polar flux axis radiates like a rod with \citep{wal84} $L_{GW} = ({32}/{45}) \mu^2 \Omega^6 \simeq ({32}/{45}) (m \Omega)^2$ with $\mu = ml^2$. A star of mass $M$, radius $R$, Newtonian binding energy $W=M^2/(2R)\simeq 0.15$ generally satisfies $M >> W >> E_{J} >> E_{turb} >> E_{B}$ for turbulent motions $E_{turb}$ and $E_B = (1/6)B^2R^3$. Hence, $E_B =  \left({E_{rot}}/{W}\right) \left({E_{turb}}/{E_{rot}}\right) \left({E_B}/{E_{turb}} \right) W \simeq 10^{-4} M$ for fiducial ratios of 0.1 for each factor with corresponding $B\simeq 4\times 10^{16}$G. $M_b$ enhances $m\simeq f_BE_B$ by $2\beta_A^{-2}$, $\beta_A=c_A/c$, where $f_B\simeq 0.5$ for a dipole field. Accordingly, $L_{GW} \simeq ({32}/{45}) \left(m\Omega\beta_A^{-2}\right)^2 \simeq 2\times 10^{52} \left( {B_{16}}/({\beta_A/0.1})\right)^4\left( {f_s}/{350\mbox{Hz}}\right)^2 \mbox{erg~s}^{-1}$ at rapid spin when $l$ is a few times $R$. Such burst will be short by canonical 
bounds on $E_J$ of a NS.

$E_J$ increases dramatically in continuing core-collapse to a BH. 
A numerical estimate of ${\cal E}$ derives from catalytic conversion of $E_{J} = 2M\sin^2(\lambda/2)\lesssim 0.29M$ at 
$a/M=\sin\lambda$ (non-extremal) at modest efficiency at orbital angular velocity $\Omega_T = \pi f_{GW}$ 
relative to $\Omega_H = \tan(\lambda/2)/(2M)$ of the BH. The estimated initial frequency of $\sim744\,$Hz at
time-of-coalescence $t_c$ inferred from $t_s=0.67\,$s is below the orbital frequency at which the stars approach 
the ISCO of the system mass, about 1100\,Hz at $r\simeq16$ km. At this point, an equal mass DNS
has $a/M = 0.72 < 1$ consistent with numerical simulations \citep[e.g.][]{bai17}, allowing collapse to a $\sim3M_\odot$ Kerr BH with 
$E_J \simeq 24\%M_\odot c^2$. For a torus radius $K$ times the ISCO radius, 
Fig. 5 shows the result of integration of the equations describing spin-down (Supplementary Data) with 
a ${\cal E} \simeq 3.6-4.3\%M_\odot$ at aforementioned canonical initial values $a/M$, subject to the observed gravitational-wave 
frequency $600\,\mbox{Hz} < f_{GW} < 700\,$Hz at $t_s=0.67$\,s post-merger - consistent with (\ref{EQN_E}).


\begin{thebibliography}{99}
\bibitem[Abbott et al.(2017a)]{abb17a} Abbott, B.P., Abbott, R., Abbott, T.D., et al., 2017a, Phys. Rev. Lett., 119, 161101
\bibitem[Abbott et al.(2017b)]{abb17b} Abbott, B.P., Abbott, R., Abbott, T.D., et al., 2017b, Astrophys. J. 848, L13
\bibitem[Abbott et al.(2017c)]{abb17c} Abbott, B.P., et al., 2017, ApJ, 851, L16
\bibitem[Ai et al.(2018)]{ai18} Ai, S., Gau, He, Dai, Z.-G., Wu, X.-F., Li, A., Zhang, B., \& Li, M.-Z., ApJ, 860, 57 (2018)
\bibitem[Baiotti \& Rezzolla(2017)]{bai17} Baiotti, L., \& Rezzolla, L. 2017, RPPh, 80, 096901
\bibitem[Bays et al.(2012)]{bay12} Bays, K., et al., 2012, Phys. Rev. D, 85, 052007
\bibitem[Connaughton(2017)]{con17} Connaughton, V. 2017, GCN, 21505
\bibitem[Coughlin \& Dietrich(2019)]{cou19} Coughlin, M. \& Dietrich, 2019, arXiv:1901.06052v1
\bibitem[Coward et al.(2002)]{cow02} Coward, D.M., van Putten, M.H.P.M., \&  Burman, R.R., 2002, ApJ, 580, 1024
\bibitem[D'Avenzo et al.(2017)]{dav17} D'Avanzo, E., Benetti, S., Branchesi, M., Brocato, E., et al, 2017, Nat., 551, 67
\bibitem[Etienne et al.(2009)]{eti09} Etienne, Z.B., Liu, Y.T.,  Shapiro, S.L., and Baumgarte, T.W., 2009, Phys. Rev. D 79, 044024
\bibitem[Faber(2009)]{fab09} Faber, J.A., 2009, Class. Quant. Grav., 26, 114004
\bibitem[Ferrari et a.(2010)]{fer10} Ferrari, V., Gualtieri, L., \& Pannarale, F., 2010, Phys. Rev. D, 81, 064026
\bibitem[Freedman(2017)]{fre17} Freedman, W. L. 2017, NatAs, 1, 0121
\bibitem[Gill et al.(2019)]{gil19} Gill, R., Nathanail, A., \& Rezzolla, L., 2019, arXiv:1901.04138v1
\bibitem[Goldstein et al.(2017)]{gol17} Goldstein, A. et al., 2017,  Astrophys. J. 848, L14
\bibitem[Guidorzi et al.(2017)]{gui17} Guidorzi, C., Margutti, R., Brout, D., Scoling, D., Fong, W., et al., 2017, ApJ, 851, L36
\bibitem[Hacyen(2017)]{hac17} Hacyan, S., 2017, Rev. Mex. Fis. 63, 466
\bibitem[Haensel et al.(2009)]{hae09} Haensel, P., Zdunik, J. L., Bejger, M., et al. 2009, A\&A, 502, 605
\bibitem[Kalapotharakos et al.(2012)]{kal12} Kalapotharakos, C., Kazanas, D., Harding, A., \& Contopoulos, I., 2012, ApJ, 749, 2
\bibitem[Kasliwal et al.(2017)]{kas17a} Kasliwal, M.M., Nakar, E., Singer, L.P., 2017, Science, 358, 1559
\bibitem[Kasen et al.(2017)]{kas17b} Kasen, D., Metzger, B., Barnes, J., et al., 2017, Nat., 551, 80
\bibitem[Kerr(1963)]{ker63} Kerr, R.P., 1963, Phys. Rev. Lett., 11, 237 
\bibitem[Lattimer \& Schramm(1976)]{lat76} Lattimer, J.M., \& Schramm, D.N., 1976, ApJ, 210, 549
\bibitem[Li et al.(2018)]{li18} Li, S.-Z., Liu, L.-D., Yu, Y.-W. \& Zhang, B., 2018, ApJ, 861, L12
\bibitem[Metzger et al.(2018)]{met18} Metzger B. D., Thompson T. A., Quataert E., 2018, ApJ, 856, 101
\bibitem[Oliver et al.(2019)]{oli19} Oliver, M., Keitel, D., Miller, A., et al., 2019, arXiv:1812.06724
\bibitem[Perego et al.(2014)]{per14} Perego, A., Rosswog, S., Cabez\'on, R.M., et al., 2014, MNRAS, 443, 3134
\bibitem[Pian et al.(2017)]{pia17} Pian, E., D'Avanzo, P.,Benetti, S., Branchesi, M., Brocato, E., et al. 2017, Nat., 551, 67
\bibitem[Piro et al.(2019)]{pir19} Piro, L., Troja, E., Zhang, B., et al., 2019, MNRAS, 483, 1912
\bibitem[Pozanenko et al.(2018)]{poz18} Pozanenko, A.S., Barkov, M.V., Minaev, P.Y., et al., 2018, ApJ, L30
\bibitem[Mooley et al.(2018a)]{moo18a} Mooley, K.P., Deller, A.T., Gottlieb, O., et al., 2018, Nat. 554, 207
\bibitem[Mooley et al.(2018b)]{moo18b} Mooley, K.P., Deller, A.T., Gottlieb, O., et al., 2018, Nat. 561, 355
\bibitem[Savchencko et al.(2017)]{sav17} Savchenko, V., Ferrigno, C. , Kuulkers, E., et al., 2017, ApJ, 848, L15
\bibitem[Siellez et al.(2013)]{sie13} Siellez, K., Bo\"er, M., \& Gendre, B., 2013, MNRAS, 437, 639 
\bibitem[Smartt et al.(2017)]{sma17} Smartt, S.J., Chen, T.-W., Jerkstrand, A., Couthlin, M., Kankare, E., et al., 2017, Nat., 551, 75
\bibitem[Vallisneri(2000)]{val00} Vallisneri, M., 2000, Phys. Rev. Lett., 84, 3519 
\bibitem[Vallisneri et al.(2014)]{val14} Vallisneri, M., et al., LOSC, Proc. 10th LISA Symp., U Florida, May 18-23, 2014; arxiv:1410.4839 
\bibitem[van Putten(1999)]{van99} van Putten, M.H.P.M., 1999, Science, 284, 115
\bibitem[van Putten(2001)]{van01} van Putten, M.H.P.M., 2001, Phys. Rev. Lett., 87, 091101
\bibitem[van Putten \& Levinson(2002)]{van02a} van Putten, M.H.P.M., \& Levinson, A., 2002, Science, 295, 1874
\bibitem[van Putten \& Levinson(2002)]{van02b} van Putten, M.H.P.M., \& Levinson, A., 2002, Class. Quant. Grav., 19, 1309
\bibitem[van Putten \& Levinson(2003)]{van03} van Putten, M.H.P.M., \& Levinson, A., 2003, ApJ,  584, 937
\bibitem[van Putten(2012)]{van12} van Putten, M.H.P.M., 2012, Prog. Theor. Phys., 127, 331
\bibitem[van Putten et al.(2014)]{van14} van Putten, M.H.P.M., Guidorzi, C.. \& Frontera, F.. 2014, ApJ, 786, 146
\bibitem[van Putten(2015)]{van15} van Putten, 2015, ApJ, 810, 7
\bibitem[van Putten \& Della Valle(2019)]{van19} van Putten, M.H.P.M. \& Della Valle, M., 2019, MNRAS, 482, L46
\bibitem[Wald(1984)]{wal84} Wald, R.M., 1984, {\em General Relativity}, University of Chicago Press
\bibitem[Yang et al.(2018)]{yan18} Yang, H., East, W.E., \& Lehner, L., 2018, ApJ, 856, 110
\bibitem[Yu et al.(2018)]{yu18} Yu Y.-W., Liu, L.-D., \& Dai, Z.-G., 2018, ApJ, 861, 114
\end{thebibliography}
\end{document}